\documentstyle[11pt,psfig]{article}

\newcommand{\mathPic}[1]{
\centerline{\leavevmode
\psfig{figure=./Figs/#1_math.ps,prolog=,height=3.25in,bbllx=0.5in,bblly=2.75in,bburx=8in,bbury=8in}}
}

\newcommand{\qed}{\vspace{.1em}\noindent\fbox{\rule{
0em}{.1em}\rule{.1em}{0em}}\vspace{1em}}

\newenvironment{algorithm}[1]{
\begin{quotation}
\small
\setlength{\baselineskip}{12pt}
\setlength{\parskip}{0pt}
\samepage
\noindent\rule{\parindent}{0.005in}\ \raisebox{-0.5ex}{\sl #1}\ \hrulefill\\
}{

\noindent\hrulefill\ 
\end{quotation}
}

\newcommand{\definitionBullet}{$\bullet$}

\newenvironment{definitions}{
\begin{list}{\definitionBullet}{}
}{
\end{list}
}

\newenvironment{proof}{

\noindent{\bf Proof:}\ }{
\hfill \qed

}

\newtheorem{theorem}{Theorem}[section]
\newtheorem{lemma}[theorem]{Lemma}
\newtheorem{corollary}[theorem]{Corollary}

\newenvironment{linearprogram}[2]{
\samepage

\begin{eqnarray*}
& & \mbox{#1\ } #2 \\
& & \mbox{subject to\ }\left\{ 
\begin{array}{rcl@{\hspace{0.2in}}l}}{
\end{array}\right.
\end{eqnarray*}
}

\newcommand{\R}{{\cal R}}
\newcommand{\N}{{\cal N}}

\newcommand{\etal}{et al.}
\newcommand{\eg}{e.g.\ }

\newcommand{\OPT}{\mbox{\sc Opt}}
\newcommand{\Opt}{\mbox{\sc Opt}}
\newcommand{\LRU}{\mbox{\sc Lru}}

\newcommand{\FIFO}{\mbox{\sc Fifo}}

\newcommand{\FWF}{\mbox{\sc Fwf}}

\newcommand{\MARK}{\mbox{\sc Mark}}
\newcommand{\Mark}{\mbox{\sc Mark}}
\newcommand{\GD}{\mbox{\sc GreedyDual}}

\newcommand{\BAL}{\mbox{\sc Balance}}
\newcommand{\Bal}{\mbox{\sc Balance}}

\newcommand{\IP}{\mbox{\sc IP}}
\newcommand{\LP}{\mbox{\sc LP}}
\newcommand{\DP}{\mbox{\sc DP}}

\newcommand{\cost}[3]{{\cal C}_{#3}({#1},{#2})}
\newcommand{\numphases}[2]{{\cal P}_{#2}({#1})}
\newcommand{\avenew}[2]{{\cal N}_{#2}({#1})}

\newcommand{\GDRelabel}{\mbox{\sc Relabel}}
\newcommand{\GDMove}{\mbox{\sc Move}}
\newcommand{\GDStay}{\mbox{\sc Stay}}

\begin{document}
\title{
The $K$-Server Dual \\ and \\ 
Loose Competitiveness for Paging}

\author{Neal Young\thanks{
    This research was performed while the author was at the
    Computer Science Department, Princeton University, Princeton, NJ
    08544, and was supported by the Hertz Foundation.
}}

\date{}

\begin{titlepage}

\maketitle

\begin{abstract}
{\em Weighted caching} is a generalization of {\em paging} in which
the cost to evict an item depends on the item.  
We give two results concerning strategies for these problems 
that incur a cost within a factor of the minimum possible on each input.

We explore the linear programming structure of the more general
{\em $k$-server} problem.
We obtain the surprising insight 
that the well-known ``least recently used'' and ``balance''
algorithms are primal-dual algorithms.
We generalize them both,
obtaining a single $\frac{k}{k-h+1}$-competitive,
primal-dual strategy for weighted caching.

We introduce {\em loose competitiveness}, motivated by Sleator and
Tarjan's complaint \cite{ST-85-2} that the standard competitive ratios
for paging strategies are too high.  A $k$-server strategy is {\em
loosely c(k)-competitive} if, for any sequence, for {\em almost all}
$k$, the cost incurred by the strategy with $k$ servers {\em either}is
no more than $c(k)$ times the minimum cost {\em or} is insignificant.
We show that $k$-competitive paging strategies
including ``least recently used'' and ``first in first out''
are loosely $c(k)$-competitive provided $c(k)/\ln k \rightarrow \infty$.
We show that the $(2\ln k)$-competitive, randomized ``marking algorithm''
of Fiat \etal\ \cite{FKLMSY-89}
is loosely $c(k)$-competitive provided $c(k)-2\ln\ln k\rightarrow \infty$.
\end{abstract}
\thispagestyle{empty}
\end{titlepage}

The body of this paper consists of four sections.  In \S 1, the
introduction, we describe background, our results, and related work.
In \S 2, we give our weighted caching strategy.  In \S 3, we show
loose competitiveness of various paging strategies.  We conclude with
comments about further research in \S 4.

\section{Introduction} 
Many real problems must be solved {\em on-line} --- decisions that
restrict possible solutions must be made before the entire problem is
known.  Generally, one can not guarantee an optimal solution if one
must solve a problem on-line.  Thus a natural question for such a
problem is whether a strategy exists that guarantees an approximately
optimal solution.

In this paper we study the {\em $k$-server} problem
\cite{mcge-85,MMS-90}.  Various definitions of the problem exist in
the literature; we take the following definition, which is technically
convenient and essentially equivalent to the other definitions: One is
given a complete directed graph with edge lengths $d(u,v)$, a number,
$k$, of identical, mobile servers, and a sequence $r$ of requests,
each to some node.  In response to the first request, all servers are
placed on the requested node.  In response to each subsequent request
$v$, if no server is on $v$, some server must be chosen to move from
its current node $u$ to $v$ at a cost of $d(u,v)$.  A strategy for
solving the problem is {\em on-line} if it chooses the server to move
independently of later requests.  The goal is to minimize the total
cost.

As many authors (\eg Chrobak and Larmore \cite{CL-91}) have pointed
out, the $k$-server problem is an abstraction of a number of practical
on-line problems, including linear search, paging, font caching, and
motion planning for two-headed disks.

We focus on two special cases of the $k$-server problem: the weighted
caching problem \cite{MMS-90}, in which $d(u,v) = w(u)$ for $u\neq v$
(the cost to move a server from a node depends only on the node), and
the paging problem \cite{ST-85-2}, in which the cost to move a server
is uniformly 1.

Traditionally, paging is described as the problem of managing a {\em
fast memory}, or {\em cache}, capable of holding $k$ items: items are
requested; if a requested item is not in the fast memory, it must be
placed in the fast memory, possibly evicting some other item to make
room.  The goal is to minimize the {\em fault rate} --- the number of
evictions per request.  Weighted caching is similar, except that the
cost to evict an item depends on the item.  

For these two problems, for technical reasons and without loss of
generality, we replace the asumption that all servers begin on the
first requested node with the assumption that initially no servers
reside on nodes, and, in response to any request, any server that has
not yet served a request may be placed on the requested node at no
cost.

Following a number of authors (Sleator and Tarjan
\cite{ST-85-2}; Borodin, Linial, and Saks \cite{BLS-87}; and Manasse,
McGeoch, and Sleator \cite{MMS-90}), we are interested in strategies
that are {\em competitive}, that is, strategies that on any sequence
incur a cost bounded by some constant times the minimum cost possible
for that sequence.  Formally,
\begin{definitions}
\item $r$ denotes an arbitrary sequence of requests.
\item $X$ denotes some on-line $k$-server strategy.
\item $k$ denotes the number of servers given to the on-line strategy.
\item \OPT\ denotes the (off-line) strategy producing a minimum cost solution.
\item $h$ denotes the number of servers given to \OPT.
\item $\cost{X}{k}{r}$ denotes the (expected) cost
incurred by the (randomized\footnote{We implicitly assume that the
input requests are independent of the random choices made by the
strategy; for other models see \cite{BBKTW-90}.}) strategy $X$ with
$k$ servers on request sequence $r$.
\item A strategy $X$ is {\em $c$-competitive} for a given $r$,
$h$, and $k$, when
\[ \cost{X}{k}{r} \le c\cdot\cost{\OPT}{h}{r} + b,\]
where $b$ depends on the initial positions of the optimal and
on-line servers, but is otherwise independent of $r$.
\item A strategy $X$ is {\em $c(h,k)$-competitive} when
$X$ is $c(h,k)$-competitive for {\em all} $r$, $h$ and $k$.  $C(h,k)$
is then called a {\em competitive ratio} of $X$.
\end{definitions}
Note that the competitiveness of a strategy is unrelated to its
computational complexity.

Before we describe our results, here is a summary of the strategies
relevant to our work.
\begin{definitions}
 \item \LRU, \FIFO, and \FWF\ are, respectively the ``least recently
used'', ``first in first out'', and ``flush when full'' paging
strategies. \LRU\ moves the server from the least recently requested,
served node. \FIFO, which can be obtained from \LRU\ by ignoring
served requests, moves the least recently moved server.  \FWF\
evicts all items from the fast memory (removes all servers from the
graph at a cost of $k$) when the fast memory is full and the requested
item is not in the fast memory.

 \item \MARK\ is the marking algorithm, a randomized paging strategy.
\MARK\ may be described as follows: if the requested node has no
server, mark all servers if none are marked, and then move and unmark
a marked server chosen uniformly at random; if the requested node has
a server, unmark that server.

 \item \BAL\ is the balance algorithm, a $k$-server strategy.  In
response to request $r$, \BAL\ moves the server from served node $u$
minimizing $d(u,r)+W(u)$, where $W(u)$ denotes the net distance
traveled so far by the server on $u$.  \BAL\ generalizes \FIFO.

\end{definitions}

\subsection{A primal-dual strategy for weighted caching}
In \S 2 we introduce and analyze \GD, a new, primal-dual,
deterministic, on-line weighted-caching strategy that is (optimally)
$\frac{k}{k-h+1}$-competitive.  Figure \ref{gdFig1} contains a direct
description of \GD.

\begin{figure}[htb]
\begin{algorithm}{\GD}

Maintain a pair of real values $L[s] \le H[s]$ with each server $s$.

In response to each request, let $v$ be the requested node;

\begin{enumerate}
\item[] {\bf if} some server $s$ is on $v$ {\bf then}
        \begin{enumerate}
        \item[] {\bf let} $H[s] \leftarrow w(v)$
        \end{enumerate}
\item[] {\bf else if} some server $s$ has yet to serve a request {\bf then}
        \begin{enumerate}
        \item[] {\bf let} $L[s] \leftarrow H[s] \leftarrow w(v)$
        \item[] Place $s$ on $v$.
        \end{enumerate}
\item[] {\bf else}
        \begin{enumerate}
        \item[\GDRelabel:]
                Uniformly lower $L[s]$ and $H[s]$ for all $s$ so that
                \[\min_{s} L[s] \le 0 \le \min_{s} H[s].\]
        \item[] Move any server $s'$ such that $L[s'] \le 0$ to $v$.
        \item[] {\bf let} $L[s'] \leftarrow H[s'] \leftarrow w(v)$
        \end{enumerate}
\end{enumerate}
\end{algorithm}
\caption{The weighted caching algorithm \GD}
\label{gdFig1}
\end{figure}

\GD\ is of practical interest because it generalizes \LRU, one of the best
paging strategies, to weighted caching.  \GD\ also generalizes \Bal\
for weighted caching, and thus \FIFO.\footnote{The natural
generalization of \LRU\ for weighted caching can be obtained by
ignoring the $L[\cdot]$ values and lowering as much as possible in the
\GDRelabel\ step.  \Bal, as it specializes for weighted caching, can be
obtained by ignoring the $H[\cdot]$ values and lowering as little as
possible in the \GDRelabel\ step.}

\GD\ is of theoretical interest because its analysis is the first
primal-dual analysis\footnote{We will assume familiarity with linear
programming primal-dual techniques.  For an introduction, see
\cite{PS-82}.} of an on-line algorithm and because the analysis, which
shows an (optimal) competitive ratio of $\frac{k}{k-h+1}$, is the
first to show a ratio less than $k$ when $h<k$ for any $k$-server
problem more general than paging.  A consequence of this reduced ratio
is that \GD\ has a constant competitive ratio provided $h$ is any
fraction of $k$.

We feel that the primal-dual approach, well developed for exact
optimization problems, is also important for approximation problems,
including on-line problems, because primal-dual considerations help
reveal combinatoric structure, especially how to bound optimal costs.
The primal-dual approach also has the potential to unify the arguably
ad hoc existing on-line analyses.  For instance, the analyses of \LRU\
and \FIFO\ \cite{ST-85-2}, of \Bal\ for weighted caching
\cite{CKPV-91}, and of \MARK\ \cite{FKLMSY-89} can all be cast as
closely related primal-dual analyses.  The primal-dual approach can
also reveal connections to existing optimization theory.  For these
reasons, we take pains to make explicit the primal-dual framework
behind our analysis.

Here is a sketch of our primal-dual approach.  The $k$-server problem
has a natural formulation as an integer linear program (IP) that is
essentially a minimum-weight matching problem.  Relaxing the
integrality constraints of IP yields a linear program (LP) (which,
incidentally, has optimal integer solutions).  \GD\ implicitly
generates a solution to the dual program (DP) of LP.  The dual
solution serves two purposes: \GD\ uses the structural information
that the solution provides about the problem instance to guide its
choices, and \GD\ uses the cost of the dual solution as a lower bound
on $\cost{\OPT}{h}{r}$ to certify competitiveness.

Related work includes the following.  Sleator and Tarjan
\cite{ST-85-2} show that \LRU\ and \FIFO\ are
$\frac{k}{k-h+1}$-competitive, and that this ratio is optimal for
deterministic, on-line paging strategies.  A similar analysis shows
that \FWF\ is also $\frac{k}{k-h+1}$-competitive.

Fiat \etal\ \cite{FKLMSY-89,youn-91-2} introduce and analyze \MARK,
showing that it is $2H_k$-competitive ($H_k \approx \ln k$), and
showing that no randomized paging strategy is better than
$H_k$-competitive when $h=k$.  McGeoch and Sleator \cite{MS-89}
subsequently give a $H_k$-competitive randomized paging strategy.
Young \cite{youn-91} shows that \MARK\ is roughly
$2\ln\frac{k}{k-h}$-competitive when $h<k$ and that no randomized
strategy is better than roughly $\ln\frac{k}{k-h}$-competitive.

Manasse, McGeoch, and Sleator \cite{MMS-90} show that \BAL\ is
$k$-competitive for the general problem provided only $k+1$ distinct
nodes are requested, and that no deterministic algorithm is better
than $\frac{k}{k-h+1}$-competitive in {\em any} graph with at least
$k+1$ distinct nodes.  

Chrobak, Karloff, Payne, and Vishwanathan \cite{CKPV-91} show that
\BAL\ is $k$-competitive for weighted caching.  Independently of their
analysis of \BAL, Chrobak, Karloff, Payne, and Vishwanathan
\cite{CKPV-91} formulate the $k$-server problem as an
integral-capacity minimum-cost maximum-flow problem and use this
formulation to give a polynomial time algorithm to find a minimum-cost
solution.

The primal-dual approach has been used extensively for exact
optimization problems \cite{PS-82}, and is used implicitly in a number
of recent analyses of approximation algorithms.  Goemans and
Williamson \cite{GW-92} explicitly use the approach for finding
approximate solutions to NP-hard connectivity problems.

\subsection{A more realistic $k$-server model}
\begin{figure}[t]
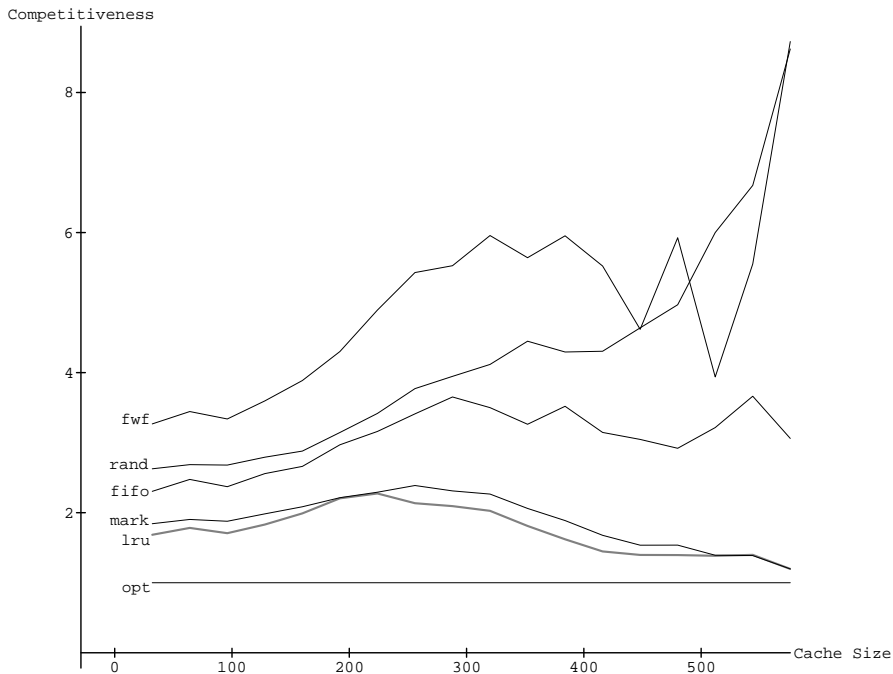

\centerline{\mathPic{competitiveness_make}}
\caption{Competitiveness 
$\left(\frac{\cost{\cdot}{k}{r}}{\cost{\OPT}{k}{r}}\right)$
vs. $k$ for typical $r$}
\label{competitivenessFig}
\end{figure}

In \S 3 we give the second contribution of this paper: {\em loose
competitiveness}.  Loose competitiveness is motivated by Sleator and
Tarjan's \cite{ST-85-2} complaint that (when $h=k$) the competitive
ratios of paging strategies are too high to be of practical interest.
We have done simulations that suggest that in practice good paging
strategies usually incur a cost within a small constant factor of
minimum.  The graph in Figure \ref{competitivenessFig} plots
competitive ratio $\cost{X}{k}{r}/\cost{\OPT}{k}{r}$ versus $k$ for a
number of paging strategies on a typical sequence.\footnote{The input
sequence, traced by Dick Sites \cite{SA-88}, consists of 692,057
requests to 642 distinct pages of 1024 bytes each.  The sequence was
generated by two X-windows network processes, a ``make'' (program
compilation), and a disk copy running concurrently.  The requests
include data reads and writes and instruction fetches.}

\begin{figure}[t]
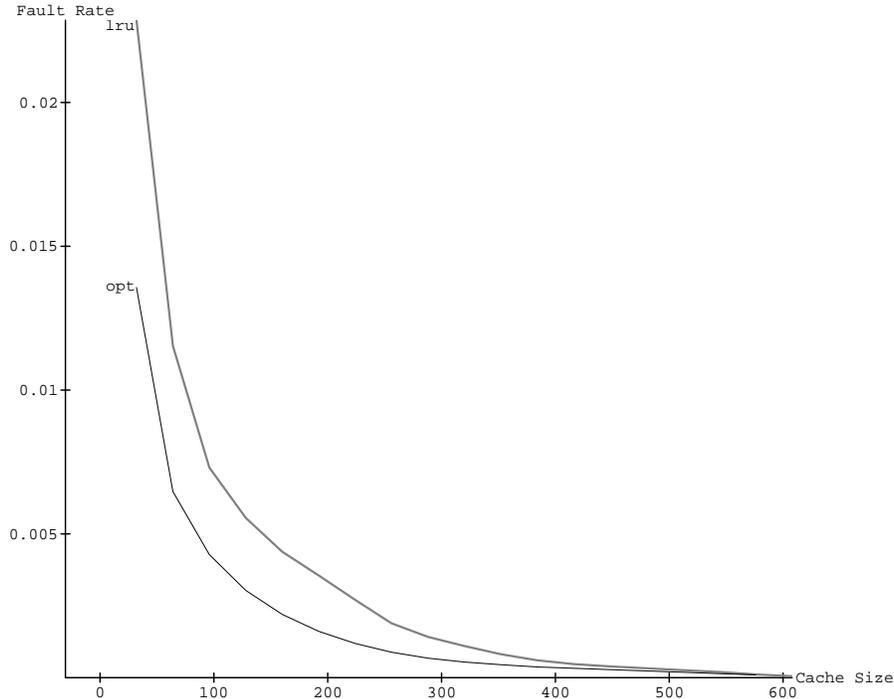

\centerline{\mathPic{faultRate_make}}
\caption{Fault rate 
$\left(\frac{\cost{\cdot}{k}{r}}{|r|}\right)$
vs. $k$ for typical $r$}
\label{faultRateFig}
\end{figure}
We would like to keep the worst-case character of competitive analysis
but somehow show more realistic competitive ratios.
\begin{definitions}
\item A strategy $X$ is {\em loosely $c(k)$-competitive} when,
for all $d > 0$, for all $n\in \N$, for any request sequence $r$, only
$o(n)$ values of $k$ in $\{1,\ldots,n\}$ satisfy
\[\cost{X}{k}{r} \ge 
\max\{c(k)\cdot\cost{\OPT}{k}{r}, \cost{\OPT}{1}{r}/n^d\}+b,\]
where $b$ depends only on the starting configurations of $X$ and \OPT,
and the $o(n)$ is independent of $r$.
\end{definitions}
That is, $X$ is loosely $c(k)$-competitive when, for any sequence, at
all but a vanishing fraction of the values of $k$ in any range
$\{1,...,n\}$, either $X$ is $c(k)$-competitive in the usual sense, or
the cost to $X$ with $k$ servers is insignificant (or both).  For
instance, if a paging strategy is loosely $3\ln\ln k$-competitive,
then, for any fixed $d>0$, on any sequence, for almost any choice of
$k$ in any range $\{1,2,...,n\}$, the fault rate will be either at
most $1/n^d$ or at most $3\ln\ln k$ times the minimum possible using a
cache of size $k$.

This model is realistic provided input sequences are not critically
correlated with $k$ and provided we are only concerned about being
near-optimal when the cost is significant.  Both criteria are arguably
true for most paging applications.

\begin{definitions}
\item A paging strategy is {\em conservative} if it moves no server
from a node until all servers have been placed on the graph, and it
moves servers at most $k$ times during any consecutive subsequence
requesting $k$ or fewer distinct items.
\end{definitions}

\LRU, \FIFO, \Mark, and even \FWF\ are conservative.  Any conservative
paging strategy is $\frac{k}{k-h+1}$-competitive \cite{youn-91-2}.

The results we obtain are as follows: any conservative paging strategy
is loosely $c(k)$-competitive provided $c(k)/\ln k
\rightarrow \infty$ and both $c(k)$ and $k/c(k)$ are non-decreasing;
\MARK\ is loosely $c(k)$-competitive provided $c(k)-2\ln\ln k
\rightarrow \infty$ and both $c(k)$ and $2\ln k - c(k)$ are
non-decreasing.

Loose competitive ratios are thus shown to be exponentially lower than
standard competitive ratios.

Borodin, Irani, Raghavan and Scheiber \cite{BIRS-91} give a related
work, in which the possible request sequences are quantified by the
degree to which they exhibit a certain kind of locality of reference,
and competitive ratios are considered as a function of this parameter.
The work is extended by Irani, Karlin, and Phillips \cite{IKP-92}.
The ratios shown in their model are, in most cases, much higher than
the loose competitive ratios established in this paper.

\section{\GD}
In this section we develop and analyse \GD.  We first develop a linear
programming framework for the general $k$-server problem, and then we
present and analyze \GD\ as a primal-dual algorithm within this framework.

\subsection{The $k$-server dual}
Fix a request sequence $r_0,r_1,...,r_N$, so that request $i$ is to
node $r_i$.

We next define \IP, an integer linear program whose feasible solutions
correspond to solutions of the $k$-server problem given by $r$.  The
variables of $\IP$ are $\{x_{ij} : 0\le i < j \le N\}$, where
$x_{ij}\in\{0,1\}$ is 1 if and only if the request served by the
server of request $j$ before serving request $j$ is request $i$.

After defining \IP, we construct its fractional relaxation \LP, and
the dual \DP\ of \LP.

\begin{definitions}
\item $\IP(k)$ (or just $\IP$, if $k$ is determined by context)
denotes the integer linear program
\begin{linearprogram}{minimize}{\sum_{0\le i < j \le N} d(r_i,r_j)x_{ij}}
x(\mbox{out}(0)) & \le & k &     \\
x(\mbox{out}(i)) & \le & 1 &            (1 \le i \le N-1) \\
x(\mbox{in}(i))  & =   & 1 &            (1 \le i \le N) \\
          x_{ij} & \in & \{0,1\} &      (0 \le i < j \le N)
\end{linearprogram}
where $\mbox{out}(i)$ denotes the set $\{(i,j) : i < j \le N\}$,
$\mbox{in}(i)$ denotes the set $\{(j,i) : 0 \le j < i\}$, and
$x(S) = \sum_{(i,j) \in S} x_{ij}$.

For the weighted caching and paging problems (where initially no
servers reside on the graph, and each server is allowed to serve its
first request by being placed on the requested node at no cost), \IP\
is defined as above, but we stipulate that request 0 is to an
artificial node that is never requested again and that is at
distance 0 to all later requests.  With this stipulation the initial
conditions for the general problem reduce to the initial conditions
for weighted caching and paging.

\item $\LP(k)$ (or just $\LP$) denotes the relaxation of \IP\ 
(obtained by replacing each constraint $x_{ij} \in \{0,1\}$ with the
constraint $0 \le x_{ij} \le 1$).

\item $\DP(h)$\ (or just \DP) denotes the dual of $\LP(h)$:
\begin{linearprogram}{maximize}{
- h a_0 - \sum_{1\le i \le N-1} a_i + \sum_{1 \le i \le N} b_i}
b_j - a_i & \le & d(i,j) & (0 \le i < j \le N)\\
      a_i & \ge & 0      & (0 \le i \le N)
\end{linearprogram}
\item $\| (a,b) \|_h$ denotes the cost, 
$-h a_0 - \sum_{i \ge 1} a_i + \sum_i b_i$,
of a feasible solution to $\DP(h)$.
\end{definitions}
Note that the dual constraints are independent of $h$, so that
a dual solution is feasible independently of $h$.  

By duality, for any feasible dual solution $(a,b)$,
\[\cost{\OPT}{h}{r} \ge \|(a,b)\|_h\]

Incidentally, a standard transformation shows that \IP\ is equivalent
to a minimum-weight, bipartite, perfect matching problem\footnote{It
may be useful for the reader in understanding \LP\ to study the
equivalent minimum-weight perfect matching problem, so we briefly
outline it here.  Construct a weighted bipartite graph $G=(U, W, E)$,
with $U=\{A_0,A_1,...,A_{N-1}\}$, $W=\{B_1,...,B_N\}$,
$E=\{(A_i,B_j)\in U\times W : i < j\}$, and $w(A_i,B_j)=d(r_i,r_j)$.
Each solution $x$ to \IP\ corresponds to the subset $\{e : x_e = 1\}$
of $E$.  The cost of $x$ equals the net weight of edges in the subset.
Such subsets are exactly those such that every vertex in $W$ touches
one edge in the subset, every vertex in $U$ except $A_0$ touches at
most one edge, and $A_0$ touches at most $k$ edges.  We can leave the
problem in this form, or we can convert it into a true perfect
matching problem by duplicating $A_0$ with its edges $k$ times and
adding $k$ copies of a new node $B_\infty$ with zero-cost edges from
every $A_i$.}, and thus that \LP\ has optimal integer solutions, so
that, for given $r$ and $h$, the above bound is tight for some
$(a,b)$.

\subsection{The algorithm}
Here are the definitions and notations specific to \GD:
\begin{definitions}
\item A request {\em has a server} if the request has been served and the
server has not subsequently served any other request.

 \item The notation $i^-$ denotes the most recent request (up to and
including request $i$) that resulted in the server of request $i$
moving.  We define $0^- = 0$.

 \item $S$ denotes the set $\{i : \mbox{ request } i \mbox{ has a
server}.\}$.  (More correctly, $S$ is a multiset, as 0 occurs in $S$
once for each server on node $r_0$.  Any $i>0$ can occur only once.)

 \item $(a,b)$ denotes a feasible dual solution maintained by \GD.

\end{definitions}

\GD\ responds to each request as follows.  If the requested node has a
server, it does nothing.  Otherwise, it uniformly raises a subset of
the dual variables enough to account for the cost of moving some
server, but not so much that feasibility is violated.  It then moves a
server whose movement cost can be accounted for.  The full algorithm
is given in Figure \ref{gdFig2}.

\begin{figure}[htb]
\begin{algorithm}{\GD(r,k)}
\item[Comment:] moves servers in response to requests $r_0,r_1,...,r_N$,
maintaining $(a,b)$, a dual solution, and $S$, a multiset containing
the currently served requests, such that
$(a,b)$ is feasible and the distance traveled by servers is at most 
\(\frac{k}{k-h+1}\|(a,b)\|_h - \sum_{i\in S} b_{i^-+1}.\)

\item[In response to request 0:] \ 
\begin{description}
\item[] {\bf let} $a_{i-1} \leftarrow b_i \leftarrow 0$ for $i=1,...,N$
\item[] {\bf let} $S \leftarrow$ the multiset containing request 0 with multiplicity $k$
\end{description}

\item[In response to each subsequent request $n>0$:] \ 

\begin{description}
\item[] {\bf if} node $r_n$ has a server {\bf then}
\begin{enumerate}
        \item[\GDStay:]
{\bf choose} $i \in S$ such that $r_i=r_n$ \\
{\bf let} $S \leftarrow S\cup \{n\}-\{i\}$, satisfying request $n$
\end{enumerate}

\item[] {\bf else}
\begin{enumerate}
\item[\GDRelabel:]
Uniformly raise the dual variables in the set
        \[\{a_i : 0 \le i \le n-1, i \not\in S\}
                \cup \{b_i : 1 \le i \le n\}\]
so that $(\forall i\in S)\ b_{i+1} \le w(r_i)$
but $(\exists i \in S)\ b_{i^-+1} \ge w(r_i)$.
\item[\GDMove:]
{\bf choose} $i \in S$ such that $b_{i^-+1} \ge w(r_i)$ \\
{\bf let} $S \leftarrow S\cup \{n\}-\{i\}$, satisfying request $n$
\end{enumerate}
\end{description}
\end{algorithm}
\caption{\GD\ as a primal-dual algorithm}
\label{gdFig2}
\end{figure}

\subsection{Analysis of the algorithm}
A simple proof by induction on $n$ shows that every $b_i \ge b_{i+1}$, that
$b_{i+1} \le w(r_i)$ for $i\in S$ (two facts that we use again later),
and that the \GDRelabel\ step can in fact be performed.  All other
steps can be seen to be well-defined by inspection, and clearly
\GD\ produces an appropriate sequence of server movements.  This
establishes the correctness of \GD.

To establish that $\GD$ is $\frac{k}{k-h+1}$-competitive, we show two
invariants: that the dual solution $(a,b)$ is feasible, and that the
distance traveled by servers is bounded by
$\frac{k}{k-h+1}\|(a,b)\|_h - \sum_{i\in S} b_{i^-+1}$.  Since
every $b_i$ is nonnegative and $\|(a,b)\|_h$ is a lower bound on
$\cost{\OPT}{h}{r}$, this gives the result.

\begin{lemma}
\GD\ maintains the invariant that $(a,b)$ is feasible.
\end{lemma}
\begin{proof}
By induction on $n$.

Clearly $(a,b)$ is initially feasible.

The only step that changes $(a,b)$ is \GDRelabel.

Clearly \GDRelabel\ maintains that every $a_i$ is nonnegative.

Thus the only dual constraint that \GDRelabel\ might violate
is of the form
\[ b_j - a_i \le d(r_i,r_j)\]
for some $0 \le i < j \le N$.

By inspection of the \GDRelabel\ step, such a constraint can only be
violated if $i \in S$ and $j \le n$.

In this case, $a_i = 0$ and $r_i \ne r_j$ because $i$ has a server,
so the constraint reduces to \(b_j \le w(r_i).\)

Since $b_{i+1} \le w(r_i)$ after the step, and $b_{i+1} \ge b_j$
(since $j>i$ and we have already established that every $b_i \ge
b_{i+1}$), the constraint is maintained.
\end{proof}

\begin{lemma} 
\GD\ maintains the invariant that the net distance traveled by servers is
bounded by
\[\frac{k}{k-h+1}\|(a,b)\|_h - \sum_{i\in S} b_{i^-+1}\]
\end{lemma}
\begin{proof}
By induction on $n$.

Clearly the invariant is initially true.

The \GDStay\ step leaves the net distance and the bound unchanged.

The \GDRelabel\ step also leaves the net distance and the bound
unchanged.  If $0 \not\in S$, that the bound remains unchanged can be
seen by inspecting the definition of $\|(a,b)\|_h$, and noting that
when the dual variables are raised, $n-k$ of the $a_i$'s, including
$a_0$, and $n$ of the $b_i$'s increase.  Consequently $\|(a,b)\|_h$ is
increased by $k-h+1$ times as much as any individual term, and, in the
bound, the increase in the minuend exactly counterbalances the
increase in the subtrahend.

If $0 \in S$, the bound remains unchanged because the constraint
$b_1\le w(r_0) = 0$ ensures that the raise is degenerate --- that the
dual variables are in fact unchanged.

The \GDMove\ step increases the distance traveled by $w(r_i)$, and
increases the bound by $b_{i^-+1} - b_{n^-+1}$.  Since $n^- = n$,
$b_{n+1} = 0$, and $b_{i^-+1} \ge w(r_i)$, the bound is increased
by at least $w(r_i)$, and the invariant is maintained.
\end{proof}

\begin{corollary}
\GD\ is $\frac{k}{k-h+1}$-competitive.
\end{corollary}

Note that in order to implement \GD, only the values $L[s_i] =
w(r_i)-b_{i+1}$ and $H[s_i] = w(r_i)-b_{i^-+1}$ (for each server $s_i$
of a request $i\in S$) need to be maintained, and that the artificial
first request may be dropped, instead placing the servers on nodes
when they first truly serve a request.  We leave it to the reader to
verify that these modifications lead to the direct description of \GD\
given in Figure \ref{gdFig1}.

\section{Loose Competitiveness}
In this section we give our analyses of loose competitiveness of
paging strategies.  The theorems and lemmas in this section, except as
noted, first appeared in \cite{youn-91,youn-91-2}.

\label{looseCompSec}
Recall that a $k$-server strategy is {\em loosely $c(k)$-competitive}
if, for any $d$, for any $n$, for any request sequence $r$, only
$o(n)$ values of $k\in\{1,...,n\}$ satisfy
\[\cost{X}{k}{r} \ge \max\{c(k) \cost{\OPT}{k}{r}, \cost{\OPT}{1}{r}/n^d\}+b.\]

The following terminology is essentially from Fiat \etal's
\cite{FKLMSY-89,youn-91-2,youn-91} analysis of the marking
algorithm.  Given a sequence $r$ and a positive integer $k$,
\begin{definitions}

 \item The {\em $k$-phases} of $r$ are defined as follows.  The first
$k$-phase is the maximum prefix of $r$ containing requests to at most
$k$ distinct nodes.  In general, the $i$th $k$-phase is the maximum
substring\footnote{By ``substring'' we mean a subsequence of
consecutive items.} of $r$ beginning with the request, if any,
following the $i-1$st $k$-phase and containing requests to at most $k$
distinct nodes.

Thus the $i+1$st $k$-phase begins with the (new) request that would
cause \FWF\ to flush its fast memory for the $i$th time.

 \item $\numphases{k}{r}$ denotes the number of $k$-phases, minus 1.

 \item A {\em new request} (for a given $k$) in a $k$-phase (other
than the first) is a request to a node that is not requested
previously in the $k$-phase or in the previous $k$-phase.  Thus in two
consecutive $k$-phases, the number of distinct nodes requested is $k$
plus the number of new requests in the second $k$-phase.

 \item $\avenew{k}{r}$ denotes the {\em average} number of new requests
per $k$-phase of $r$ other than the first.

Thus the total number of new requests in $r$ for a given $k$ is
$\avenew{k}{r}\cdot\numphases{k}{r}$.

\end{definitions}

Our analysis has two parts.  In the first part (Theorem
\ref{looseThm}) we show that, for any sequence, few values of $k$
yield both a large number of $k$-phases and a low average number of
new requests per $k$-phase.

In the second part, we show (Lemma \ref{lcstep1}) that, for the paging
strategies that interested us, for a given sequence and $k$, the cost
incurred by the strategy is proportional to the number of $k$-phases,
and the competitiveness is inversely related to the average number of
new requests per $k$-phase.  Consequently (Corollary \ref{looseCor}),
by the first part of the analysis, few values of $k$ yield both a high
cost and a high competitiveness.

The key technical insight (Lemma \ref{looseLemma}) for the first part
of the analysis is that if, for a given $k$, the average number of new
requests per $k$-phase is low, then, for $k'$ just slightly larger
than $k$, the number of $k'$-phases is a fraction of the number of
$k$-phases.
\begin{lemma}
\label{looseLemma}
Fix a sequence $r$.  For any $k$, and any $k' \ge k+2\avenew{k}{r}$,
\[\numphases{k'}{r}\le\frac{3}{4}\numphases{k}{r}.\]
\end{lemma}
\begin{proof}
Let $p_0,\ldots,p_{\numphases{k}{r}}$ denote the $k$-phase
partitioning of $r$.

At least half (and thus at least $\lceil \numphases{k}{r}/2 \rceil$) of the
$\numphases{k}{r}$ $k$-phases $p_1,...,p_{\numphases{k}{r}}$ have a number
of new requests not exceeding $2\avenew{k}{r}$.  Denote these by 
\(p_{i_1},\ldots,p_{i_{\lceil \numphases{k}{r}/2 \rceil}}.\)

If we modify the $k$-phase partitioning of $r$ by joining
$p_{i_j-1}$ and $p_{i_j}$ for odd $j$, we obtain a coarser
partitioning of $r$ into at most $\numphases{k}{r}-\lceil \numphases{k}{r}/4
\rceil$ pieces.  In the coarser partitioning, each piece resulting from a
join references at most $k+2\avenew{k}{r} \le k'$ distinct nodes,
while each pieces remaining unchanged from the $k$-phase partioning
references at most $k \le k'$ distinct nodes.

If we now consider the $k'$-phase partitioning, we find that each
$k'$-phase must contain the final request of at least one of the
pieces in the coarser partition, because if a $k'$-phase begins at or
after the beginning of a subsequence of requests to at most
$k+2\avenew{k}{r}$ distinct nodes, it will continue at least through
the end of the subsequence.

Thus $\numphases{k'}{r}
\le \numphases{k}{r} - \lceil\numphases{k}{r}/4\rceil 
\le \frac{3}{4} \numphases{k}{r}$.
\end{proof}

From this we show that there are not too many values of $k$ yielding
both a low average number of new requests per $k$-phase and a
significant number of $k$-phases:
\begin{theorem}
\label{looseThm}
For any $\epsilon > 0, M > 0$, and any sequence $r$, the number of $k$
satisfying
\begin{equation}
\label{looseCond}
\avenew{k}{r} \le M \mbox{\ \ and\ \ } 
\numphases{k}{r} \ge \epsilon \numphases{1}{r}
\end{equation}
is $O(M \ln \frac{1}{\epsilon})$.
\end{theorem}
\begin{proof}

Let $s$ be the number of $k$ satisfying the condition.

We can choose $l=\left\lceil s/\lceil 2M\rceil \right\rceil$ such $k$
so that each chosen $k$ differs from every other by at least $2M$.
Then we have
\(1\le k_1\le k_2 \le\ldots\le k_l\) such that for each $i$
\begin{eqnarray}
\avenew{k_i}{r} & \le & M, \label{L1} \\
k_{i+1} - k_i & \ge & 2M, {\rm\ and} \label{L2} \\
\numphases{k_l}{r} & \ge & \epsilon \numphases{1}{r}. \label{L3}
\end{eqnarray}

Then for any $i$, by (\ref{L1}) and (\ref{L2}), $k_{i+1}\ge
k_i+2\avenew{k_i}{r}$, so, by Lemma \ref{looseLemma}, $\numphases{k_{i+1}}{r}
\le (3/4)\numphases{k_i}{r}$.  Inductively, $\numphases{k_l}{r} \le (3/4)^{l-1}
\numphases{1}{r}$.

This, and (\ref{L3}), imply $(3/4)^{l-1} \ge \epsilon$, so
\[\left\lceil s/\lceil 2M\rceil \right\rceil - 1
 = l-1 \le \ln_{4/3} \frac{1}{\epsilon}.\]
This implies the bound on $s$.
\end{proof}

This establishes the first part of the analysis.  

We begin the second part by showing that \OPT's cost per $k$-phase is
at least proportional to the average number of new requests per $k$-phase:
\begin{lemma}
\label{optBoundLemma}
For arbitrary paging request sequence $r$, and arbitrary $k,h > 0$,
\begin{eqnarray}
\label{optBound}
\cost{\Opt}{h}{r}/\numphases{k}{r} & \ge & (k-h+\avenew{k}{r})/2
\end{eqnarray}
\end{lemma}
(We use only the case $k=h$, but prove the general case.)
\begin{proof}
Let $m_i$ ($1<i\le \numphases{k}{r}$) denote the number of new
requests in the $i$th $k$-phase, so that
$\avenew{k}{r}\,\numphases{k}{r} = \sum_{i>1} m_i$.  During the
$i-1$st and $i$th $k$-phases, $k+m_i$ distinct nodes are referenced.
Consequently, any strategy for $r$ with $h$ servers makes at least
$k+m_i-h$ server movements during the two phases.  Thus the total cost
for the strategy is at least
\[\max\left\{\sum_{i\ge 1} (k-h+m_{2i+1}),\sum_{i\ge 1} (k-h+m_{2i})\right\}
\ge (k-h+\avenew{k}{r})\numphases{k}{r}/2.\]
\end{proof}

We next show that the strategies that interest us incur a cost
proportional to $k$ times the number of $k$-phases, and (using the
above lemma) that the strategies have competitiveness inversely
related to the average number of new requests per $k$-phase:

\begin{lemma}
\label{lcstep1}
Let $X$ denote any conservative paging strategy.  Let \MARK\ denote
the marking algorithm. Then
\begin{eqnarray}
\numphases{k}{r} & \ge & \cost{X}{k}{r}/k \label{phasesBound} \\
\avenew{k}{r} & \le & 2k\frac{\cost{\Opt}{k}{r}}{\cost{X}{k}{r}}
\label{xBound} \\
\avenew{k}{r} & \le & 
k\exp\left(1-\frac{1}{2}\frac{\cost{\Mark}{k}{r}}{\cost{\Opt}{k}{r}}\right)
\label{markBound}
\end{eqnarray}
\end{lemma}
\begin{proof}
Bound (\ref{phasesBound}) follows directly from the definition of
conservativeness.

Bound (\ref{xBound}) follows from Bound (\ref{phasesBound}) and Bound
(\ref{optBound}) of Lemma \ref{optBoundLemma}, applied with $h=k$.

Finally, we prove Bound (\ref{markBound}).  Fix a request sequence
$r$, and let $m_i$ $(1<i\le \numphases{k}{r})$ denote the number of
new requests in the $i$th $k$-phase.  Fiat \etal\ \cite{FKLMSY-89}
show\footnote{
\MARK\ moves a server chosen uniformly at random from those on nodes
not yet requested in the current $k$-phase.

Briefly, the analysis of \MARK\ classifies nonnew requests within a
phase into {\em repeat} requests (to nodes already requested this
phase) and {\em old} requests (to nodes requested in the previous
phase but not yet in this phase); the expected cost for the $i$th old
request is bounded by $m/(k-i+1)$ because at least $k-m-i+1$ of the
$k-i+1$ nodes requested in the previous phase but not in this phase
are served, each with equal probability.}
that \(\cost{\Mark}{k}{r} \le \sum_i m_i(H_k - H_{m_i}+1)\).

Since
\(H_a - H_b = \sum_{i=a+1}^b \frac{1}{i} \le \int_a^b \frac{1}{x} dx =
\ln \frac{a}{b}\), letting $f(m) = m(1+\ln\frac{k}{m})$,
\(\cost{\Mark}{k}{r} \le \sum_i f(m_i).\)
Since $f$ is convex, 
\(\cost{\Mark}{k}{r} \le \numphases{k}{r}\,f(\avenew{k}{r}).\)
Applying Bound (\ref{optBoundLemma}),
\(\cost{\Mark}{k}{r} \le 2\cost{\Opt}{k}{r}\,f(\avenew{k}{r})/\avenew{k}{r},\)
which is equivalent to Bound {\ref{markBound}}.
\end{proof}

We have established (Theorem \ref{looseThm}) that for any sequence
there are few values of $k$ yielding many $k$-phases and a low average
number of new requests per $k$-phase.

We have established (Lemma \ref{lcstep1}) that for the strategies we
are interested in, the cost they incur with $k$ servers on a sequence
is proportional to the number of $k$-phases, while the competitiveness
is inversely related to the average number of new requests per $k$-phase.

Finally, we combine the two parts to show that, for any sequence,
there are few values of $k$ for which our strategies incur high cost
and high competitiveness.  This establishes loose competitiveness.

\begin{corollary}
\label{looseCor}
Let $X$ denote any conservative paging strategy and
$C:\N^+\rightarrow\R^+$ a nondecreasing function.

$X$ is loosely $c(k)$-competitive provided that $k/c(k)$ is
nondecreasing and
\begin{equation}
\label{xAssumption}
\frac{c(k)}{\ln k} \rightarrow \infty,
\end{equation}
while \Mark\ is loosely $c(k)$-competitive provided $2\ln k-c(k)$ is
nondecreasing and
\begin{equation}
\label{markAssumption}
c(k) - 2\ln\ln k \rightarrow \infty.
\end{equation}
\end{corollary}
\begin{proof}
Let $X$ denote either any conservative paging strategy, in which case we
assume condition (\ref{xAssumption}) and that $k/c(k)$ is nondecreasing, or
\Mark, in which case we instead assume condition
(\ref{markAssumption}) and that $2\ln k - c(k)$ is nondecreasing.

We show that, for any $d>0$, $n>0$, and request sequence $r$, the number of
{\em violators} $k\in\{1,\ldots,n\}$ is $o(n)$, where a violator is a $k$ such
that
\[\cost{X}{k}{r} \ge \max\{c(k)\cost{\Opt}{k}{r},\cost{\Opt}{1}{r}/n^d\}.\]

Let $k$ be a violator.  Then bound (\ref{phasesBound}) implies
\begin{equation}
\label{vEqn1}
\numphases{k}{r} \ge \frac{\cost{X}{k}{r}}{k} 
        \ge \frac{\cost{\Opt}{1}{r}}{n^{d+1}}
        = \frac{1}{n^{d+1}}\numphases{1}{r}.
\end{equation}

Bound (\ref{xBound}) and the monotonicity of $k/c(k)$ imply
\begin{equation}
\label{vEqn2}
\avenew{k}{r} \le 2k\frac{\cost{\Opt}{k}{r}}{\cost{X}{k}{r}}
        \le \frac{2k}{c(k)}
        \le \frac{2n}{c(n)}.
\end{equation}
Since each violator $k$ satisfies (\ref{vEqn1}) and (\ref{vEqn2}),
by Theorem \ref{looseThm}, the number of violators is
$O\left(\left(\ln n^{d+1}\right)n/c(n)\right)$.  This is $o(n)$
by assumption (\ref{xAssumption}).

If $X=\Mark$, then bound (\ref{markBound}) and the monotonicity of 
\(2\ln k-c(k)\) imply, for each violator $k$, that
\begin{eqnarray}
\avenew{k}{r} & \le & 
k\exp\left(1-\frac{1}{2}\frac{\cost{\Mark}{k}{r}}{\cost{\Opt}{k}{r}}\right)
\nonumber \\
       & \le & k\exp(1-c(k)/2) \nonumber \\
       & \le & n\exp(1 - c(n)/2),\label{vEqn3}
\end{eqnarray}
so that by bounds (\ref{vEqn1}) and (\ref{vEqn3}) and Theorem
\ref{looseThm} the number of violators is
$O\left(\left(\ln n^{d+1}\right)n\exp(1 - c(n)/2)\right)$.  
This is $o(n)$ by assumption (\ref{markAssumption}).

\end{proof}

\section{Concluding Remarks}
We conclude in this section with comments about further avenues of
research.

Historically, the role of duality in solving optimization problems is
well-explored: dual solutions are used to guide the construction of
primal solutions and to certify optimality.  For on-line problems such
as the $k$-server problem, duality can serve a similar role; the
differences are that the solutions we seek are approximate, and that
the problem we want to solve is on-line.  For on-line problems, it
seems natural to seek a sequence of {\em closely related} dual
solutions, one for each prefix of the request sequence.

For those interested in extending our approach to the general
$k$-server problem we give the following brief hints.  Solutions with
monotonic $b_i$'s are not sufficient to give good bounds: add
constraints $b_{i+1} \le b_i$ to the dual problem and reformulate the
primal; in the new primal request sequences can be much cheaper than
in the old.  Raising {\em all} of the $b_i$'s is probably not a good
idea: consider a request sequence with requests from two infinitely
separate metric spaces; a $b_i$ should change only when a request is
made to the metric space of $r_i$.  Finally, a promising experimental
approach: if $r$ is a {\em worst-case} sequence for a $k$-competitive
algorithm $X$, and the bound $\cost{X}{k}{r} \le k \|(a,b)\|_k$ is
sufficient to establish competitiveness, then $(a,b)$ must be optimal;
thus by examining {\em optimal} dual solutions for worst-case
sequences, we may discover the special properties of the (generally
non-optimal) dual solutions that we seek for such an analysis.  A
similar technique has been tried for potential functions, but in that
case each experiment is much less informative: it reveals only a
single number, not an entire dual solution.

There is a suggestive similarity between potential function and
primal-dual techniques \cite{youn-91-2}.  Briefly, both can be viewed
as transforming the costs associated with operations so that a sum of
local inequalities gives the necessary global bound.  This connection
might yield some insight into the special nature of primal-dual
analyses for on-line problems.

Open questions remain concerning loose competitiveness for paging.  In
\cite{youn-91-2}, Theorem \ref{looseThm} is shown to be tight, and
consequently the analysis of loose competitiveness for \FWF\ is shown
to be tight.  No lower bounds on the loose competitive ratios of \LRU,
\FIFO, or \Mark\ have been shown.

Finally, two challenges: find a randomized algorithm for weighted
caching that is better than $k$-competitive, and show reduced loose
competitiveness for a weighted-caching algorithm.  A possible hint:
the concept of ``new requests'' used in analyzing \MARK\ and showing
loose competitiveness of paging strategies may be captured by an
algorithm that mimics \GD, but increases each $b_i$ at only half the
rate that \GD\ does, and increases each $a_i$ only as much as
necessary to maintain the dual constraints.

\bibliographystyle{alpha}
\bibliography{competitive,linearProgramming,all}
\end{document}